\documentclass[iop]{emulateapj}
\accepted{to AJ {December 6, 2014}}
\usepackage{natbib}
\usepackage{graphicx}
\usepackage{mathtools}
\usepackage{subfigure}

\newcommand{\cotwo}{\mbox{\rm CO\,(2\,--\,1)}}
\newcommand{\coone}{\mbox{\rm CO\,(1\,--\,0)}}
\newcommand{\co}{\mbox{\rm CO}}
\newcommand{\hi}{\mbox{\rm H$\,$\scshape{i}}}

\newcommand{\kmpers}{\mbox{km~s$^{-1}$}}

\shorttitle{Single-Dish vs Interferometric \co\ Velocity Dispersions}
\shortauthors{Cald\'{u}-Primo et al.}
\slugcomment{}

\begin{document}

\title{Spatially extended and high\,--\,velocity dispersion molecular component in spiral galaxies:\\[2pt]
Single-dish vs. Interferometric Observations}

\author{
Anahi Cald\'{u}-Primo  \altaffilmark{1}
Andreas Schruba \altaffilmark{2}
Fabian Walter \altaffilmark{1}
Adam Leroy \altaffilmark{3}
Alberto D.\,Bolatto \altaffilmark{4}
Stuart Vogel \altaffilmark{4}}

\altaffiltext{1}{Max-Planck-Institut f\"ur Astronomie, K\"onigstuhl 17, 69117 Heidelberg, Germany; caldu@mpia.de}
\altaffiltext{2}{Max-Planck-Institut f\"ur Extraterrestrische Physik, Giessenbachstr. 1, 85748 Garching, Germany}
\altaffiltext{3}{National Radio Astronomy Observatory, 520 Edgemont Road Charlottesville, VA 22903, USA}
\altaffiltext{4}{Department of Astronomy, University of Maryland, College Park, MD 20742-2421, USA}

\begin{abstract}
Recent studies of the molecular medium in nearby galaxies have provided mounting evidence that the molecular gas can exist in two phases: one that is clumpy and organized as molecular clouds and another one that is more diffuse. This last component has a higher velocity dispersion than the clumpy one. In order to investigate these two molecular components further, we compare the fluxes and line widths of CO in NGC\,4736 and NGC\,5055, two nearby spiral galaxies for which high\,--\,quality interferometric as well as single\,--\,dish data sets are available. Our analysis leads to two main results: 1) Employing three different methods, we determine the flux recovery of the interferometer as compared to the single\,--\,dish to be within a range of 35\,--\,74\% for NGC\,4736 and 81\,--\,92\% for NGC\,5055, and 2)  when focusing on high (SNR$\ge$5) lines of sight, the single\,--\,dish line widths are larger by $\sim$\,(40\,$\pm$\,20)\% than the ones derived from interferometric data; which is  in agreement with stacking all lines of sight. These results point to a molecular gas component that is distributed over spatial scales larger than 30$^{\prime\prime}$($\sim$\,1\,kpc), and is therefore filtered out by the interferometer. The available observations do not allow us to distinguish between a truly diffuse gas morphology and a uniform distribution of small clouds that are separated by less than the synthesized beam size ($\sim$\,3$^{\prime\prime}$  or $\sim$\,100\,pc), as they would both be invisible for the interferometer. This high velocity dispersion component has a dispersion similar to what is found in the atomic medium, as traced through observations of the \hi\ line. 

\end{abstract}


\section{Introduction}
\label{intro}
In the classical picture of the interstellar medium (ISM), molecular gas is thought to be organized mostly inside giant molecular clouds (GMCs), typically traced by CO emission, with temperatures of $10-20$ K and internal velocity dispersions of $\sim 2-8$\,\kmpers\ \citep[e.g.,][]{bo08}. This simple picture, however, does not hold at the Galactic level, were different studies have shown that diffuse CO emission is pervasive within our Galaxy \citep[and references therein]{po88,li98,go08}. Based on studies of molecular gas in our Galaxy, molecular clouds are generally classified as either \emph{diffuse} (A$_{V}\,\sim$\,0.2, 30\,$\lesssim$\,T(K)\,$\lesssim$\,100), \emph{translucent} (A$_{V}\,\sim$\,1\,--\,2, 15\,$\lesssim$\,T(K)\,$\lesssim$\,50), or \emph{dense} (A$_{V}\,\sim$\,5\,--10, 10\,$\lesssim$\,T(K)\,$\lesssim$\,50) \citep[e.g.,][]{sn06}. On the one hand, studying the ISM in the Galaxy might seem advantageous from the sensitivity and resolution points of view. Galactic studies can carry out observations of many different molecules, isotopologues, and transitions; while extragalactic studies typically detect only the most abundant molecule in its brightest transition.  On the other hand, due to projection effects and uncertainties in distance determinations, it is difficult to quantify the amount of gas present in this diffuse molecular component within our Galaxy. Nearby galaxies offer the opportunity to study this suggestive additional molecular gas component from an outside perspective. 

Molecular gas is preferentially observed through CO emission, which can be mapped both with single\,--\,dish telescopes and with interferometers. Single\,--\,dish telescopes, especially after the development of multi-beam heterodyne receiver arrays, provide a large instantaneous field of view  and good sensitivity, although at limited spatial resolution. Interferometers, on the other hand, provide higher spatial resolution, with the drawback that not all spatial scales can be recovered. The largest spatial scale to which an interferometer is sensitive to, is determined by the shortest baselines (i.e., the closest projected distance between two individual telescopes in the plane perpendicular to the line of sight). Thus, any emission arising from structures larger than this spatial scale is `invisible' to the interferometer. If most gas inside galaxies were distributed in structures (e.g., molecular clouds) that are smaller than this largest spatial scale (and that are separated from each other by at least the synthesized beam size), then either instrument would detect the same gas component, only with different spatial resolution. Therefore, the flux recovered by the interferometer depends on the gas distribution and the molecular cloud spatial scales present in the observed galaxy as compared to the specific resolution of the interferometer's configuration.\footnote{During the image deconvolution the flux recovery will also strongly depend on the noise properties \citep[e.g.,\,][]{he02}.}

\citet{pe13} conducted a study of the molecular gas phase as traced by the  \coone\ emission in M51. They used single\,--\,dish IRAM 30m data, Plateau de Bure Interferometer (PdBI) data, and a combination of both data sets, to compare flux recovery and velocity dispersions measured in these three data sets. They show that the interferometer recovers only (50\,$\pm$\,10)\% of the total flux, indicative of a molecular component that is missed by the interferometer. In addition, they found that the purely single\,--\,dish velocity dispersions are typically twice as large as the interferometric velocity dispersions. As stated before, molecular gas might be present in different cloud types, with different temperatures, densities, and spatial distribution. These different physical configurations would modify the expected velocity dispersions present in the gas. Thus, their result implies not only that $\sim$\,50\% of the gas is missed by the interferometer, but that this gas gives rise to larger velocity dispersions compared to the velocity dispersions measured for the molecular gas (i.e.,\,CO emission) inside the GMC sample (which the interferometer is sensitive to).

In a previous paper, \citet{cp13}  investigated the differences in velocity dispersion of atomic (\hi) gas \citep[from the THINGS survey,][]{wa08} and of molecular gas \citep[from the HERACLES CO survey,][]{le09} in a sample of 12 nearby spiral galaxies. On spatial scales of $\sim$\,1\,kpc we measured similar velocity dispersions for both \hi\ and CO (as measured by a single\,--\,dish telescope), and interpreted this as the existence of a high\,--\,dispersion molecular gas component.

Based on these two previous results, we deem important to rigorously investigate the presence of the molecular gas component giving rise to these large velocity dispersions and determine its mass fraction as compared to the molecular gas inside molecular clouds. The existence of such a molecular gas component would require that CO single\,--\,dish and interferometric observations be analyzed and interpreted using different assumptions about the origin of the emission. For example, studies of the relationship of molecular gas and recent star formation rate in both nearby or distant galaxies \citep[e.g., review by][]{ke12} implicitly assume that molecular gas as traced by CO observations is inside molecular clouds and participates in the current star formation process; whereas the assumption of a significant diffuse molecular component would allow for some of the observed molecular gas being not directly (or at all) related to the current star formation activity.
In this study we will further investigate the presence of a wide\,--\,spread, high\,--\,velocity dispersion molecular gas component by comparing CO line widths as measured from interferometric and single\,--\,dish observations, as well as the fluxes measured by means of both instruments. Unfortunately, high sensitivity interferometric maps of nearby galaxies are still very scarce, a situation that will however soon change with the advent of ALMA. Thus, we concentrate our current study on two galaxies: NGC\,4736 and NGC\,5055. We compare single\,--\,dish CO data from the HERACLES Survey \citep{le09} and the ``Nobeyama CO Atlas of Nearby Spiral Galaxies'' \citep{ku07} to interferometric observations taken by \citet{la10} using the Combined Array for Research in Millimeterwave Astronomy (CARMA).  

The paper is organized as follows: In Section~\ref{data} we discuss the main characteristics of the employed data sets and the two galaxies studied. In Section~\ref{met} we discuss our methodology, to finally show the results in Section~\ref{res}. A discussion and summary is provided in Section~\ref{sum}.

\vspace{2mm}
\section{Data and Sample}
\label{data}

In order to carry out the analysis, the data need to meet three conditions: high sensitivity, a clear contrast between the scales recovered by the interferometer and by the single\,--\,dish, and available \hi\, velocity maps. Even though there is a large survey, BIMA SONG \citep{hel03}, in which the \coone\ molecular emission in nearby galaxies has been mapped with an interferometer, its sensitivity and resolution render it not suitable for this specific study. Therefore, we limit our study to the 2 galaxies for which high\,--\,sensitivity and resolution CARMA interferometric maps exist.

\subsection{\coone\ Data}
\subsubsection{Interferometric Data}
The interferometric CO data come from the CARMA interferometer. Both galaxies are mapped in the \coone\ line covering the inner $\sim 3^\prime$\ diameter of the targets via a 19\,--\,point mosaic. Observations were taken between January 2007 and July 2008 using the C, D, and E configurations resulting in a synthesized beam size of $3.51\,\times\,2.97^{\prime\prime}$ for NGC\,4736 and of $3.24\,\times\,2.81^{\prime\prime}$ for NGC\,5055. The correlator was setup with two overlapping 62\,MHz spectral windows to yield a spectral resolution of 1.95\,MHz (or 5.08\,\kmpers\ at 115 GHz)\footnote{The now publicly available data for NGC\,4736 has 10~\kmpers\, since the original data were unfortunately lost, and cannot be recovered.}.

The CARMA data were calibrated using MIRIAD \citep{sa95}, weighted by noise variance, and applied robust weighting. The data were cleaned down to a  cut\,--\,off of 1.5 times the theoretical rms noise in each pixel using MOSSDI2, which performs a steer CLEAN on a mosaicked image \citep{st84}. The images are primary\,--\,beam corrected. The flux scale was determined by observations of flux standards, including Uranus and Mars. In \citet{la10} a full description of the data reduction can be found. In their analysis they compare the CARMA observations with the BIMA SONG data \citep{hel03}. After verifying the flux scales of both datasets they concluded that no flux rescaling was required.  The rms noise sensitivity at 10\,\kmpers\ resolution is 27\,mJy\,beam$^{-1}$ (237.5\,mK)\footnote{All sensitivities in K are stated using the beam size of the  original data.} for NGC\,4736 and 19\,mJy\,beam$^{-1}$ (193.2\,mK) for NGC\,5055.

\subsubsection{Single\,--\,dish Data}
We also use single\,--\,dish \coone\ mapping from the ``Nobeyama CO Atlas of Nearby Spiral Galaxies'' \citep{ku07}, which encompasses 40 nearby spiral galaxies at distances smaller than 25\,Mpc. The observations were carried out with the Nobeyama 45m telescope and typically cover most of the optical disk with a spatial resolution of $15^{\prime\prime}$. 

In the case of NGC\,4736, the receiver used was BEARS, which consists of 25 beams \citep{su00}. In this case digital spectrometers were used as backends. The total bandwidth and frequency resolution were 512\,MHz and 605\,kHz, which at 115\,GHz correspond to 1331\,\kmpers\ and 1.57\,\kmpers, respectively.

In the case of NGC\,5055, a 4\,--\,beam SIS receiver was used, together with acousto\,--\,optical spectrometers (AOS) used as backends. This configuration yields a frequency resolution and channel spacing of 250\,kHz and 125\,kHz, respectively, providing a total bandwidth of 250\,MHz. At 115\,GHz this corresponds to velocity resolution and velocity coverage of 0.65\,\kmpers\ and 650\,\kmpers, respectively.

In both cases the spectra were then smoothed to 5\,\kmpers\ \citep{na94} as part of the data reduction. This is the spectral resolution of the data we use. At 10\,\kmpers\ resolution, the rms noise sensitivity is 69\,mJy\,beam$^{-1}$ (28.3\,mK) for NGC\,4736 and 172\,mJy\,beam$^{-1}$ (70.7\,mK)  for NGC\,5055.

\subsection{\cotwo\ Single\,--\,dish Data}
We use \cotwo\ data from the HERACLES survey \citep{le09}.  This survey used the IRAM 30m telescope to map molecular gas from 48 nearby galaxies inside the optical radius (r$_{25}$). This survey used the HERA (a 9\,--\,beam dual\,--\,polarization heterodyne receiver array) together with the Wideband Line Multiple Autocorrelator (WILMA) backend. WILMA consists of 18 units of 2\,MHz channel width each and 930\,MHz bandwidth, yielding a velocity resolution of 2.6\,\kmpers\ in a 1200\,\kmpers\ velocity range at 230\,GHz. The spatial resolution obtained at this frequency is 13$^{\prime\prime}$ after gridding. The rms noise sensitivity at 10\,\kmpers\ is of 90\,mJy\,beam$^{-1}$ (11.5\,mK) for NGC\,4736 and 94\,mJy\,beam$^{-1}$ (12\,mK) for NGC\,5055.

\subsection{\hi\ Data}
For our analysis we use \hi\ intensity weighted velocity maps. These maps come from the THINGS survey \citep{wa08}. This survey encompasses 34 nearby galaxies at distances of 2$-$15 Mpc. The observations  for the two galaxies we analyze have high spectral (5.2 \kmpers) and spatial ($\sim$11$^{\prime\prime}$ for natural weighting) resolutions.

\subsection{Sample}
\label{sample}
As mentioned in the introduction, our sample is limited by the availability of sensitive, wide-field interferometric data sets. We thus focus on two galaxies only, NGC\,4736 and NGC\,5055. They have the following properties:\\

\emph{NGC\,4736} is an isolated spiral galaxy of type SAab \citep{ke03}. It is at a distance of $5.20\,\pm\,0.43$\,Mpc \citep{to01}, has an inclination of 41$^\circ$ \citep{bl08}, and an optical radius R$_{25} \approx 5$\,arcmin ($\sim 7.5$\,kpc) \citep{ke03}. 

\emph{NGC\,5055} is a flocculent spiral galaxy classified as SAbc \citep{ke03}. Its distance is $7.8\,\pm\,2.3$\,Mpc \citep{ma05}, its inclination is 59$^\circ$ \citep{bl08}, and its optical radius is R$_{25} \approx 6$\,arcmin ($\sim 13$\,kpc) \citep{ke03}. 

\begin{figure*}[htb]
\centering
\epsscale{2.2}\plottwo{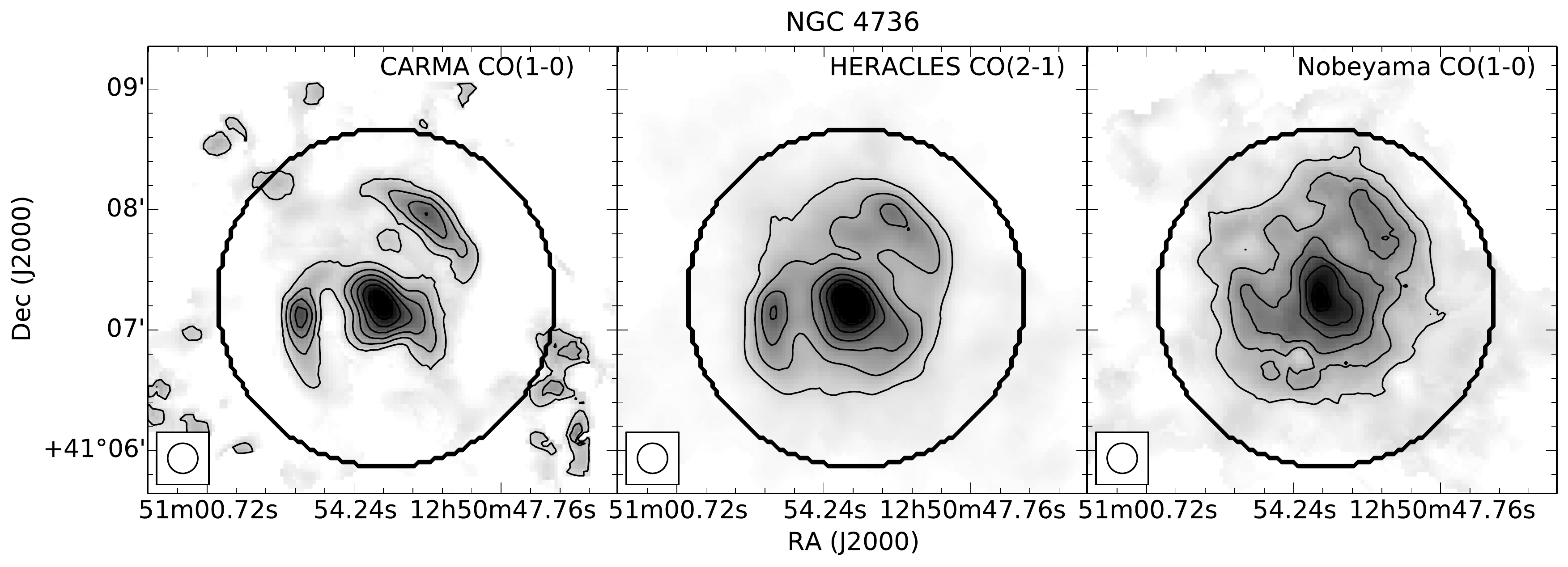}{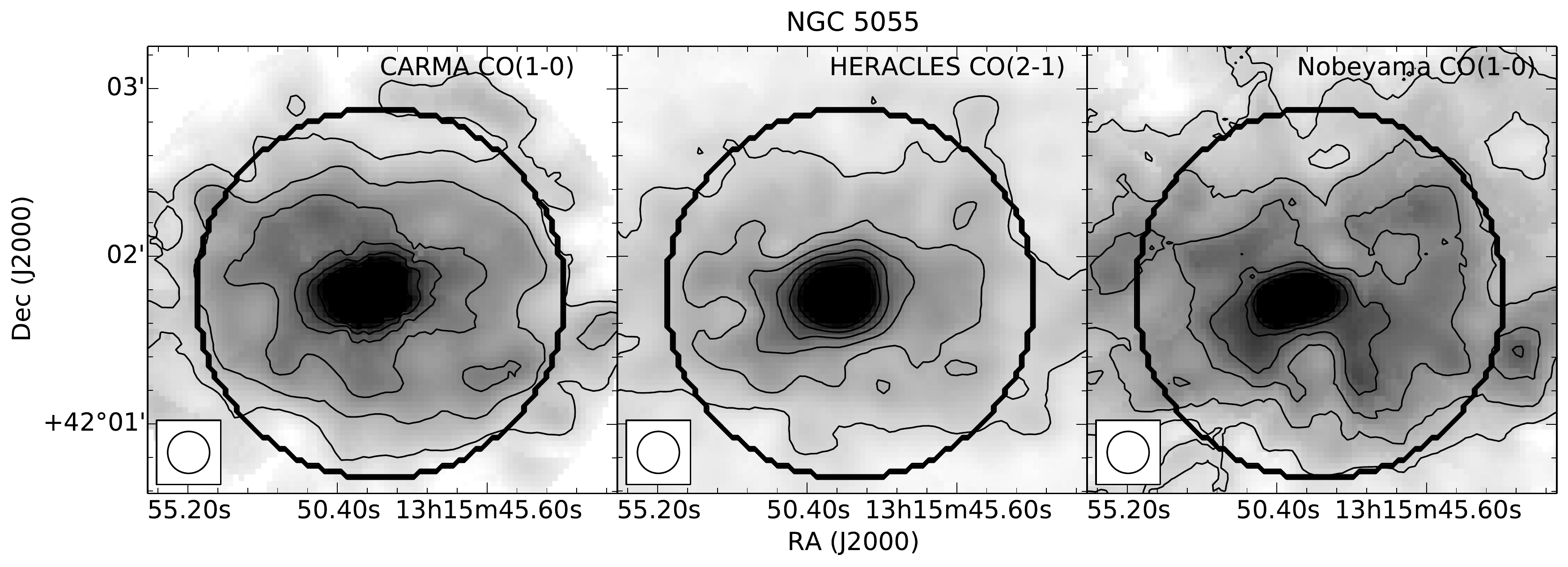}
\caption{CO integrated intensity maps of NGC\,4736 (top) and NGC\,5055 (bottom) at $15^{\prime\prime}$ resolution (the beam size is shown at the lower left corner). The first column shows the CARMA \coone\ interferometric data, the second column shows the HERACLES IRAM 30m \cotwo\ maps, and the third column shows the Nobeyama 45m \coone\ maps. The three maps were constructed using the HERACLES mask (details in Section\,\ref{flcomp}). For both galaxies the intensity range goes from 0 to 38 K\,\kmpers, and the 6 contours show intensity levels separated by 6 K\,\kmpers, starting at 6 K\,\kmpers. The black circle shows the region where the CARMA cube has homogeneous noise properties,\,i.e.,\,where the rms is within 30\% of the value at the center of the map.}\label{fig:fig1}
\end{figure*}

More infomation on the galaxy properties can be found in \citet{ke03}, \citet{wa08}, and \citet{le09}.
We show the integrated intensity CO maps of the three different surveys at a common resolution of $15^{\prime\prime}$  in Figure~\ref{fig:fig1}.

\section{Methodology} 
\label{met}

\subsection{Data Homogenization}
For our analysis we convolve the data to match the spatial and spectral resolution in all cases. This common resolution is defined by the poorest resolution of our data sets: the spectral resolution is set by the CARMA data, and the spatial resolution is set by the Nobeyama data. After hanning smoothing the data to a common spectral resolution (which is discussed below), we use the \texttt{REGRID} task in MIRIAD \citep{sa95} to put all the data sets on a common grid.  Finally, we convolve the data sets to the $15^{\prime\prime}$ Nobeyama limiting spatial resolution. 

The limiting spectral resolution is different for the two galaxies. NGC\,4736 has a CARMA\,--\,limited 10\,\kmpers\ spectral resolution, while NGC\,5055 has a CARMA/Nobeyama\,--\,limited 5\,\kmpers\ spectral resolution. Therefore, for NGC\,5055 we first carry out the analysis at a 5\,\kmpers\ spectral resolution and then compare the results when using a 10\,\kmpers\ spectral resolution.  We find that our results do not change. Thus, for sake of uniformity, we here present the analysis using a common spectral resolution of 10\,\kmpers\ for both galaxies.

In order to correctly Nyquist sample the data, we construct a hexagonal grid with a 7.5$^{\prime\prime}$ spacing (half of the beam size).

\begin{figure*}[htb]
\centering
\epsscale{1.1}\plotone{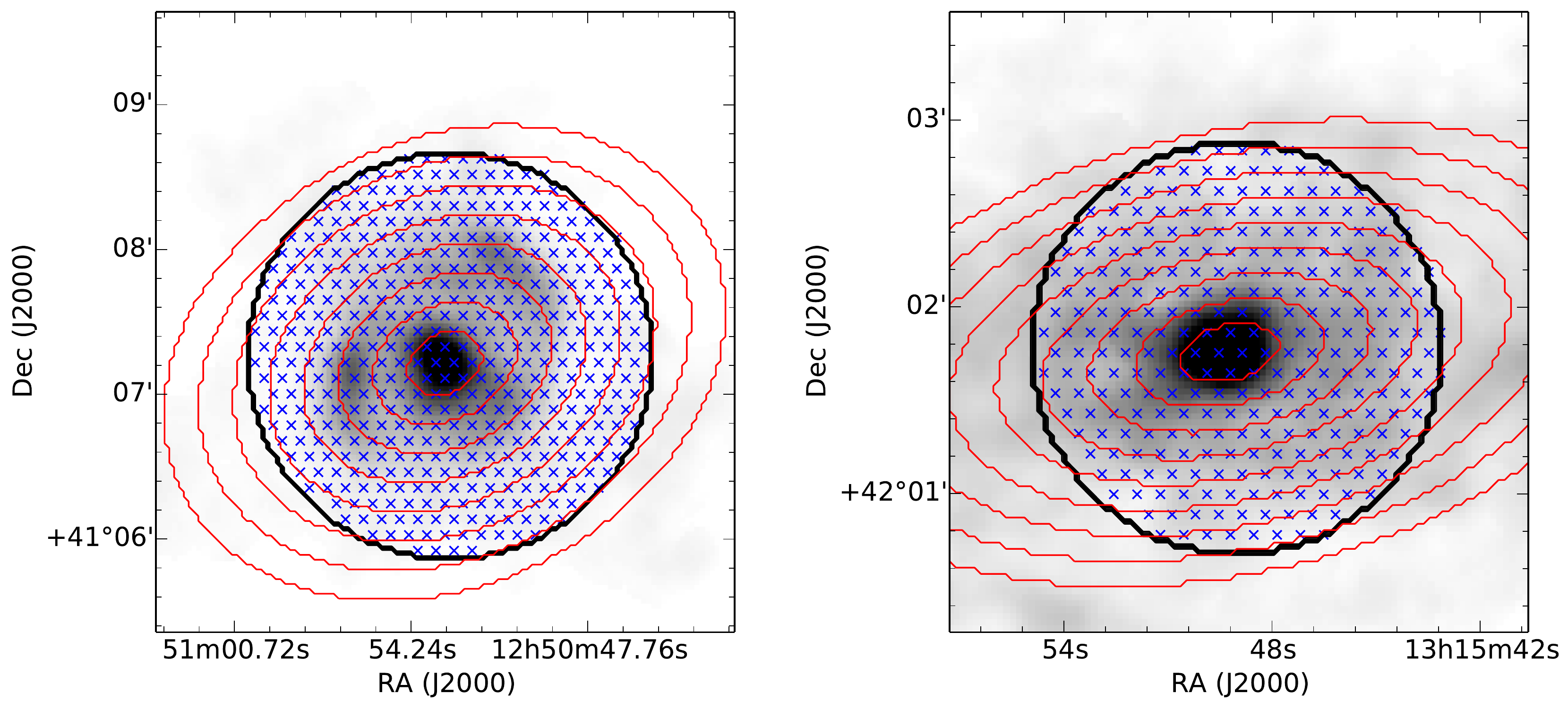}
\caption{HERACLES IRAM 30M CO integrated intensity maps of NGC\,4736 (left) and NGC\,5055 (right) at $15^{\prime\prime}$ resolution (grayscale). The black solid circle represents the area where the interferometric  primary beam sensitivity is within 30\% of the value of that at the center of the map. The blue points inside the circle show the hexagonal grid points that are used for sampling the data. The red ellipses show the 15$^{\prime\prime}$ radial bins used for stacking. From inside out and in units of r$_{25}$ they are, for NGC\,4736: 0.04, 0.10, 0.17, 0.23, 0.29, 0.35, 0.41, and 0.46; and for NGC\,5055: 0.03, 0.07, 0.11, 0.15, 0.19, 0.23, 0.27, and 0.30.}\label{fig:fig2}
\end{figure*}

Due to the attenuation of the primary beam in interferometric observations, the sensitivity drops towards the edges of the map. We construct a sensitivity mask (after primary\,--\, beam correction of the cube) from the interferometric noise map and apply it to all data sets. In this way we make sure: 1) to use the same individual lines\,--\,of\,--sight (LOS) in all data sets, and 2) that we only use LOS which have homogeneous noise properties in the interferometric data. To do so, we derive the galactocentric distance at which the noise in the CARMA observations is increased by 30\%, and only use this region for further analysis. For clarification, we show this region as a solid black circle in Figure\,\ref{fig:fig2} for each galaxy on top of the HERACLES IRAM 30M integrated intensity map in gray shades. Inside the circle we also show our hexagonal grid points (blue x symbols). Finally, we also show the 15$^{\prime\prime}$ radial bins used later in this paper as red ellipses.

From Figure\,\ref{fig:fig2} we note that the sensitive region defined by the CARMA primary beam response is independent of the inclination of the galaxy. However, the radial bins are ellipses because of the inclination. This results in having lower covering factors within the ellipses for the outermost radial bins. To quantify this effect, we present in Table \ref{tab1}\, the number of LOS from the hexagonal grid that fall within each radial bin and within our mask (third column), as compared to the number of LOS that would be present without taking into account the sensitivity mask (second column).

\begin{deluxetable*}{c c c  c c c c} 
\tablecolumns{7}
\tablewidth{\textwidth}
\tablecaption{Filling factors within each 15$^{\prime\prime}$ radial bin\label{tab1}}
\tablehead{\colhead{} & \colhead{\# grid points } & \colhead{\# grid points in} & \multicolumn{4}{c}{\# of significant grid points, SNR\,$>$\,5 \tablenotemark{b}} \\
\colhead{R$_{25}$}  & \colhead{in radial bin} & \colhead{ CARMA\,--\,sens. region (\%)\tablenotemark{a}} &\colhead{CARMA (\%)} & \colhead{HERACLES (\%)} & \colhead{Nobeyama (\%)} &\colhead{THINGS (\%)} \\
\colhead{1} & \colhead{2} & \colhead{3} & \colhead{4} & \colhead{5} & \colhead{6} & \colhead{7} \\ 
\hline \\ 
\multicolumn{7}{c}{NGC\,4736}}
\startdata
0.04 & 10 & \phantom{0}10\phantom{000}(100) & \phantom{0}4\phantom{00}(40) & \phantom{0}9\phantom{000}(90) & \phantom{0}3\phantom{00}(30) & \phantom{00}6\phantom{000}(60) \\
0.10 & 35 & \phantom{0}35\phantom{000}(100) & 28\phantom{00}(80) & 32\phantom{000}(91) & 20\phantom{00}(57) & \phantom{0}32\phantom{000}(91) \\
0.17 & 57 & \phantom{0}57\phantom{000}(100) &  44\phantom{00}(77) & 57\phantom{00}(100) & 40\phantom{00}(70) &\phantom{0}57\phantom{00}(100) \\
0.23 & 73 & \phantom{0}73\phantom{000}(100) & 34\phantom{00}(47) & 73\phantom{00}(100) & 43\phantom{00}(59) & \phantom{0}73\phantom{00}(100) \\
0.29 & 101 & 101\phantom{000}(100) & 12\phantom{00}(12) & 72\phantom{000}(71) & 24\phantom{00}(24) & 101\phantom{00}(100) \\
0.35 & 116 & \phantom{0}99\phantom{0000}(85) & \phantom{0}0\phantom{000}(0) & 43\phantom{000}(43) & \phantom{0}0\phantom{000}(0) & \phantom{0}99\phantom{00}(100) \\
0.41 & 142 & \phantom{0}64\phantom{0000}(45) &  \phantom{0}1\phantom{000}(2) & 25\phantom{000}(39) & \phantom{0}0\phantom{000}(0) & \phantom{0}61\phantom{000}(95) \\
0.46 & 165 & \phantom{0}10\phantom{00000}(6) &   \phantom{0}0\phantom{000}(0) & \phantom{0}3\phantom{000}(30) & \phantom{0}0\phantom{000}(0) & \phantom{00}8\phantom{000}(80)\\
\cutinhead{NGC\,5055}
0.03 & 9 & \phantom{0}9\phantom{000}(100) & \phantom{0}6\phantom{00}(67) & \phantom{0}9\phantom{00}(100) & \phantom{0}4\phantom{00}(44) & \phantom{0}9\phantom{00}(100) \\
0.07 & 23 & 23\phantom{000}(100) & \phantom{0}7\phantom{00}(30) & 20\phantom{000}(87) & 16\phantom{00}(70) & 20\phantom{000}(87) \\
0.11 & 34 & 34\phantom{000}(100) & \phantom{0}4\phantom{00}(12) & 34\phantom{00}(100) & 33\phantom{00}(97) & 34\phantom{00}(100) \\
0.15 & 54 & 54\phantom{000}(100) &  12\phantom{00}(22) & 54\phantom{00}(100) & 47\phantom{00}(87) & 54\phantom{00}(100) \\
0.19 & 66 & 54\phantom{0000}(82) &  11\phantom{00}(20) & 54\phantom{00}(100) & 48\phantom{00}(89) & 54\phantom{00}(100) \\
0.23 & 86 & 44\phantom{0000}(51) & \phantom{0}9\phantom{00}(20) & 39\phantom{000}(89) & 27\phantom{00}(61) & 39\phantom{000}(89) \\
0.27 & 89 & 27\phantom{0000}(30) & \phantom{0}5\phantom{00}(19) & 27\phantom{00}(100) & 18\phantom{00}(67) & 27\phantom{00}(100) \\
0.30 & 117 & 26\phantom{0000}(22) &  11\phantom{00}(42) & 24\phantom{000}(92) & 14\phantom{00}(54) & 26\phantom{00}(100)
\enddata
\tablenotetext{a}{Number of grid points within the CARMA sensitive region in a particular bin. In parenthesis the is the  \% of these grid points compared to the total number of grid points within each radial bin (i.e. compared to Column 2).}
\tablenotetext{b}{For each instrument and for each radial bin we show the number of LOS with SNR\,$>$\,5 and within the CARMA\,--\,sensitive region (black circle in Fig.\,\ref{fig:fig1}). In parenthesis is the \% of these points compared to Column 3.}
\end{deluxetable*} 
\vspace{5mm}

\subsection{Measuring Line Widths for Individual LOS} \label{LOS}
We estimate the spectral line widths using single Gaussian fits. In order to obtain reliable measurements for individual  lines of sight (LOS), we require that they have high significance. Therefore, we begin by computing the line widths for the LOS which comply with the following characteristics:  \emph{a)}\,lie within the CARMA sensitive region (black circle in Fig.\,\ref{fig:fig1}), \emph{b)}\,have peak SNR\,$> 5$\footnote{This SNR is computed in the 10\,\kmpers spectral resolution and 15$^{\prime\prime}$ spatial resolution cubes.}, and \emph{c)}\,have integrated intensities of at least 10$\sigma$, where $\sigma=\Delta\,v\sqrt{n}\,rms$ and \emph{n} is the number of measured points used to fit the line. For each of these high SNR LOS we use the \hi\ first moment value,\,i.e., the local intensity\,--\,weighted velocity, as a proxy for the velocity where we expect to find the corresponding CO line (We take the \hi\ data from the THINGS survey \citep{wa08}). This is based on our earlier findings that the \hi\, and CO line\,--\,of\,--\,sight velocities agree well at the resolution we are working at \citep{sch11,cp13}. We attempt to fit a single Gaussian to each spectrum using \texttt{MPFIT}  (IDL procedure from Craig Markwardt) inside a window of 100\,\kmpers\, centered in the \hi\, first moment value. If the fitted Gaussian is broader than 100\,\kmpers, we proceed to refit with a double\,--\,horn profile (a Gaussian scaled by a symmetric second order polynomial \citet{sai07}). Double\,--\,horn profiles are mostly encountered in the inner regions of galaxies, where bulk motions or beam smearing could be dominant. These will be discarded for our analysis (see below). The integrated intensity selection criteria (b)  removes LOS with significant negative structures (`bowls') in the interferometric data\footnote{These negative bowls are expected to be present in interferometric data because of the extrapolation required to overcome the incompleteness of the u\,--\,v coverage.}.

The errors are computed as discussed in detail in \citet{cp13}: for each LOS we add random noise to the original spectrum and redo the line width measurement. We repeat this procedure a 1000 times. The adopted error in the line width measurement is the dispersion between the results of these 1000 iterations.

\subsection{Measuring Line Widths for Stacked Spectra}\label{st}
The main purpose of the stacking is to increase the SNR in our data as well as to include all available LOS in our analysis. The stacked spectra are the most meaningful measurements, as they represent the luminosity\,--\,weighted FWHM including \emph{all} LOS inside a tilted ring (and inside the CARMA sensitive region). In order to stack individual LOS coherently, we first remove the circular motions of the four data sets using the \hi\ first moment value. The THINGS survey has very high sensitivity, resulting in reliable velocity maps. Even in the case of NGC\,4736, which has low content of \hi\ in the central parts, the peak SNR is larger than 5$\sigma$ in 97.3\% of the LOS inside 0.5\,R$_{25}$. Once all spectra are centered in the same velocity frame (in this case zero velocity) they can be stacked coherently. The stacking procedure, together with the error determination, are described in detail in \citet{cp13}. After the stacking is done, we compute the line widths as discussed above. The total error in the determination of line widths has contributions both due to noise (as modeled for the individual LOS above) as well as due to the specific selection of spectra that have been stacked \citep[which we assess through bootstrapping following][]{cp13}.

\section{Results}
\label{res}

\subsection{FWHM as function of Galactocentric Distance}
\label{r1}
\begin{figure*}[htb]
\epsscale{2}\plottwo{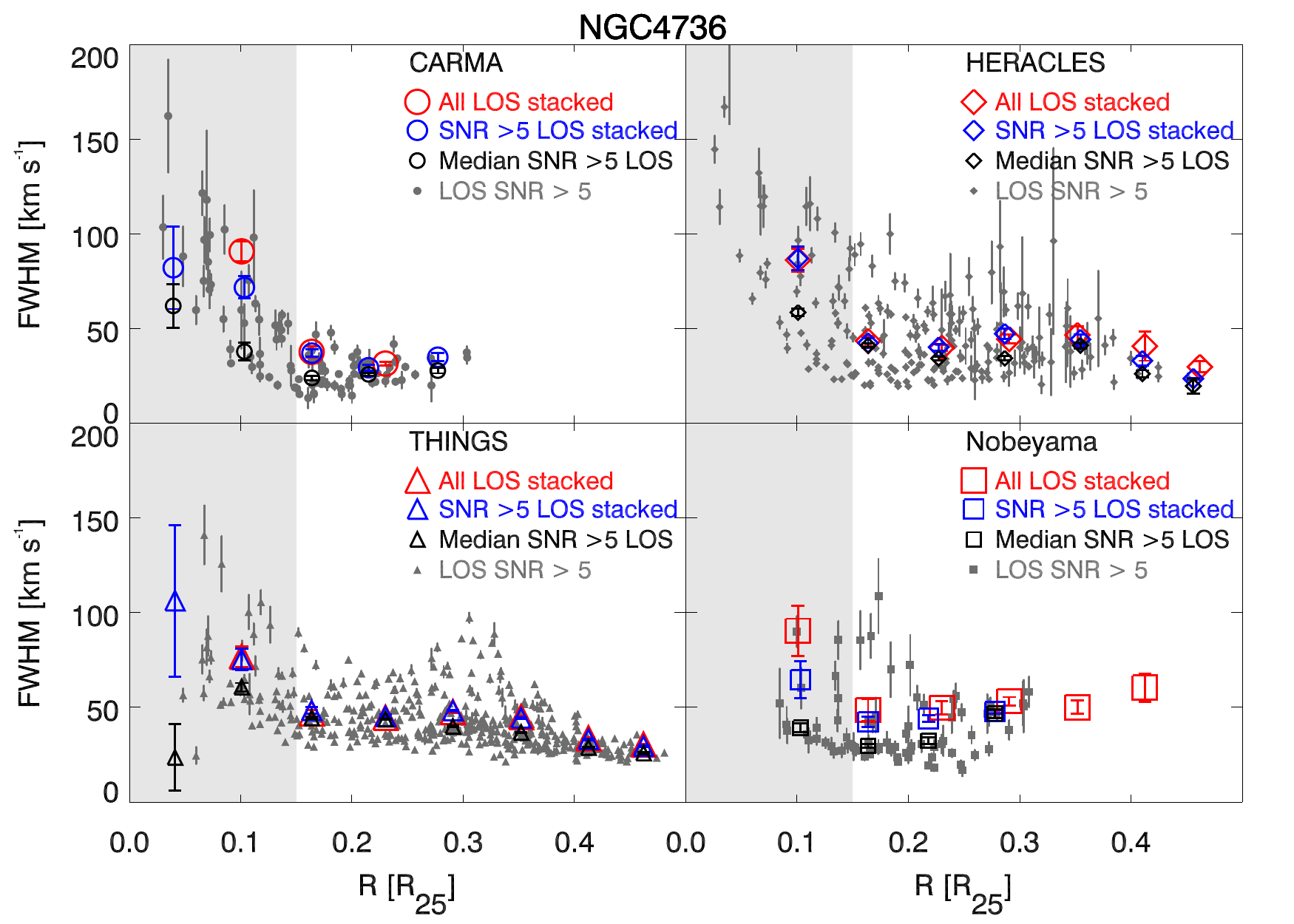}{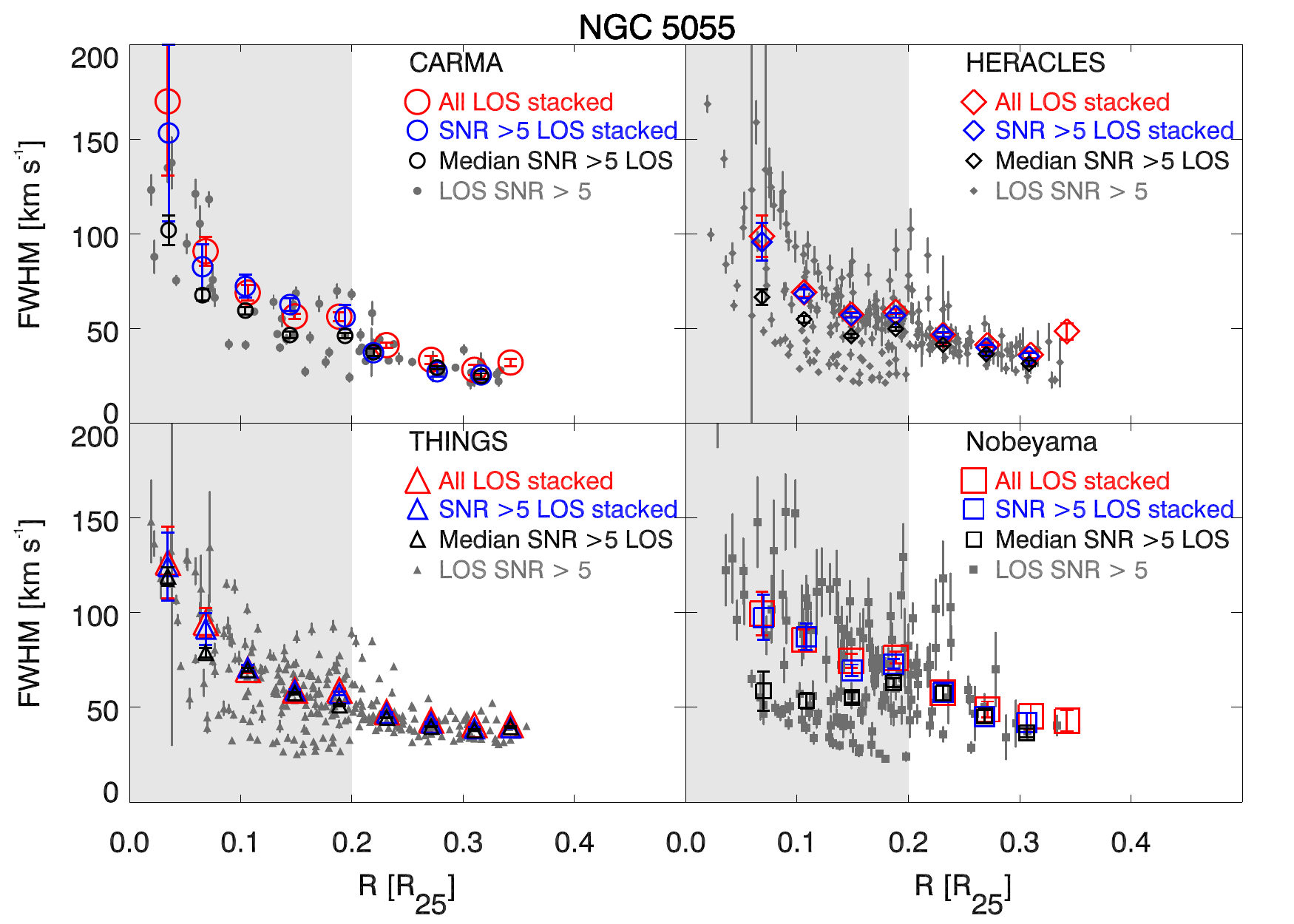}
\caption{The measured FWHM as a function of galactocentric distance for the four different instruments: CARMA (top left), HERACLES (top right), THINGS (bottom left), and Nobeyama (bottom right). The small grey symbols show individual LOS with peak SNR\,$>$\,5. The large red symbols show the values obtained when stacking all LOS within a given tilted ring bin, the medium blue symbols show the results when stacking spectra with SNR\,$>$\,5 only, and the black small symbols show the median values for the grey points,\,i.e.,\,the median values of the individual high SNR LOS measurements.  Inside the shaded area beam smearing contributes by more than 30\% to the measured line width.}
\label{fig:fig3}
\end{figure*}

We can select which LOS we want to stack. Since many galactic properties depend on the distance to the center of the galaxy, we choose to stack the individual LOS inside tilted rings of $15^{\prime\prime}$ width within each galaxy. As a first step, we stack only individual LOS with SNR\,$>$\,5. We then compare the measurements obtained from these (high SNR) stacked spectra, to the median values measured from the individual high SNR LOS. We do this comparison to test the validity of the stacking method. In Figure\,\ref{fig:fig3} we plot the FWHM measurements as function of galactocentric distance for each of the 4 different data sets (we include \hi\ for sake of completeness). The FWHM measured when fitting high significance (SNR\,$> 5$) individual LOS are shown in small grey symbols. The FWHM measured when only stacking the grey points (i.e., individual LOS with SNR\,$>5$) are shown in medium\,--\,sized blue symbols, whereas the median FWHM of the individual high SNR LOS are shown in black small symbols (The red points are explained below).
We see that at small radii (R\,$\lesssim$\,0.2\,R$_{25}$) we observe a rapid increase in the measured FWHM, however, here the observed spectra are significantly broadened by beam smearing \citep[see][for a thorough discussion on this effect]{cp13}. The shaded regions highlight the radial range where beam smearing broadens the intrinsic line profile by more than 30\%. \citet{cp13} show that for a galaxy with characteristics similar to NGC\,4736 beam smearing accounts for $>$\,30\% of the measured line width at R\,$\lesssim$\,0.15\,R$_{25}$ and for more than 10\% out to 0.25\,R$_{25}$. In NGC\,5055, beam smearing accounts for $>$\,30\% of the observed line widths at R\,$\lesssim$\,0.2\,R$_{25}$, and $>$\,10\% at 0.4\,R$_{25}$. At large radii out to 0.5\,R$_{25}$, where line profiles are only mildly or not at all affected by beam smearing, we find roughly constant CO velocity dispersions, as have been found already in earlier studies \citep[e.g.,][] {ta09, cp13}. In the regions of interest (where beam\,--\,smearing is not dominant), the measurements obtained for LOS with SNR\,$>$\,5 are in agreement.

In Table\,\ref{tab1}\, (columns 4\,--\,7)  we show the number of LOS with SNR larger than 5 within each $15^{\prime\prime}$ radial bin and within the CARMA\,--\,sensitive region for each of the datasets (in parenthesis we show which percentage of the total number of LOS within the CARMA\,--\,sensitive region these high SNR LOS represent). It is important to note that we select the individual LOS with high SNR independently for each data set,\,i.e.,\,a high SNR LOS in the HERACLES cube does not necessarily have such a high SNR LOS  counterpart in the CARMA cube\footnote{This problem does not exist when stacking all LOS. In that case we do not perform a preselection, we simply stack all LOS falling within the corresponding radial bin.}(see Table\,\ref{tab1}). It is clear that by taking only high SNR measurements we are excluding a large percentage of LOS from the analysis.

\begin{figure*}[htb]
\centering
\subfiguretopcaptrue
\subfigure[hang,nooneline][NGC\,4736]
{\epsscale{1.}\plotone {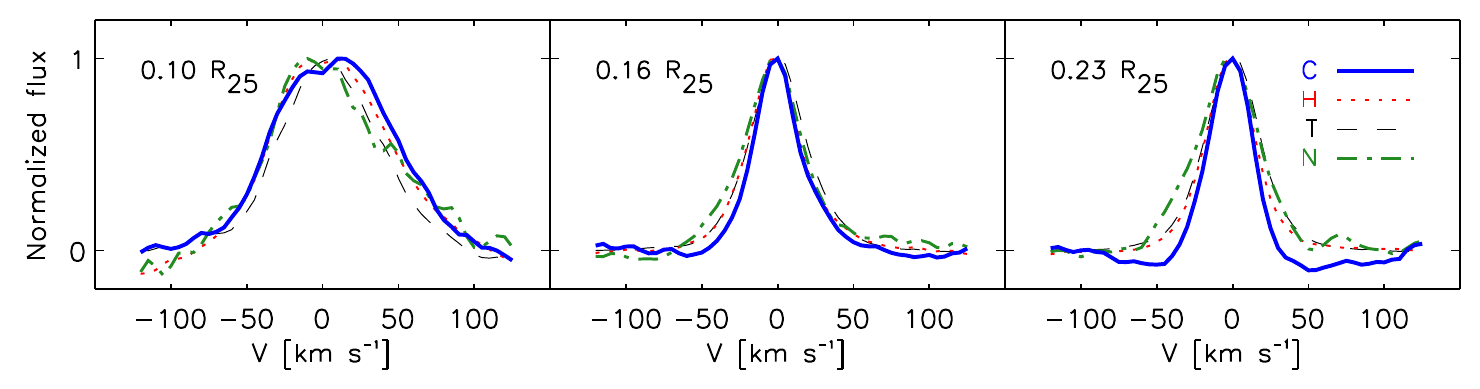}}
\subfigure[hang,nooneline][NGC\,5055 ]{\epsscale{1}\plotone{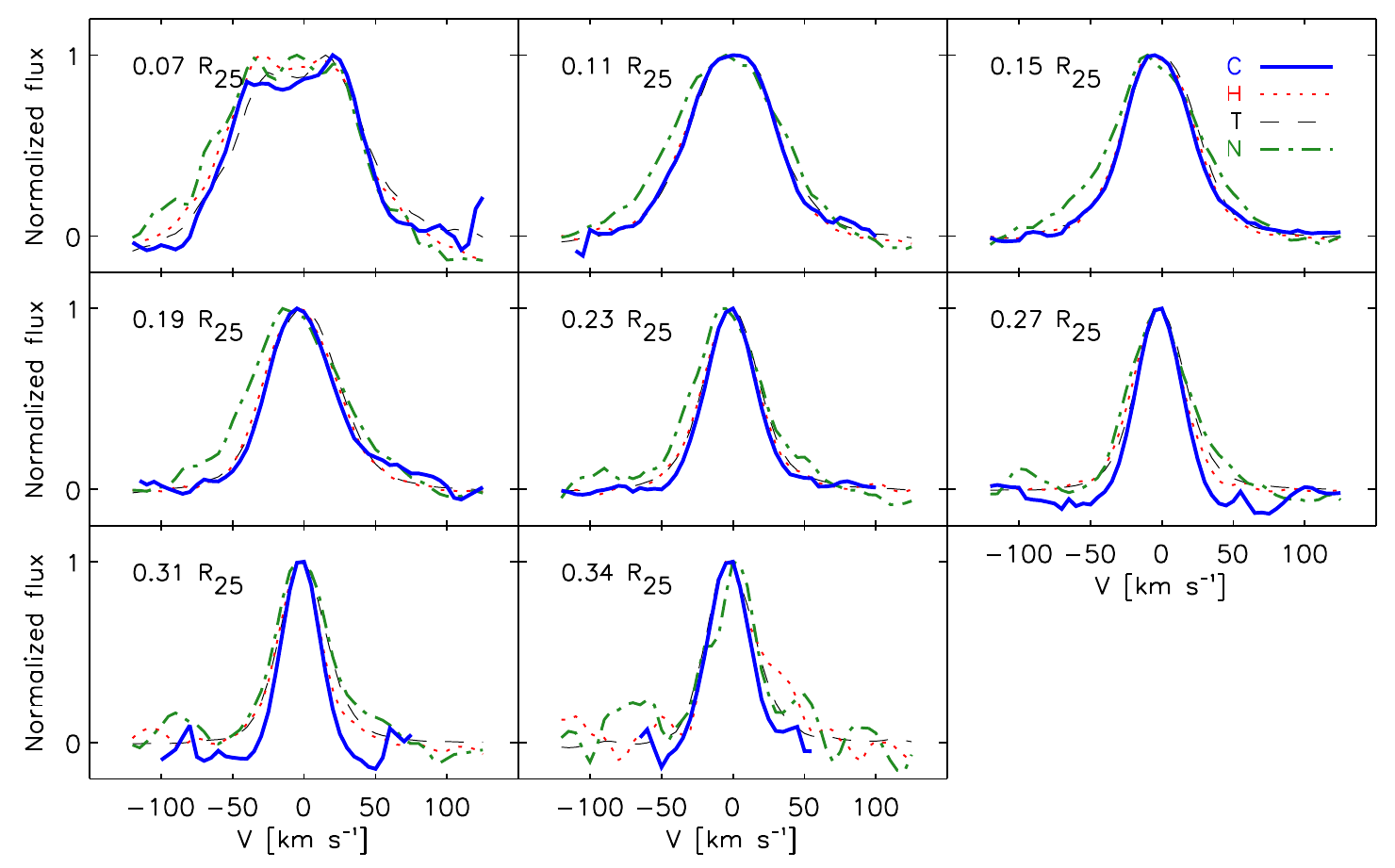}}
\caption{Stacked spectra by galactocentric distance in bins of $15^{\prime\prime}$ width. The spectra for the 4 data sets are shown together, therefore fluxes have been rescaled to peak intensities of unity. The black line (labeled `C') is the CARMA data, the red dashed line (`H') is the HERACLES data, and the green dash-dotted line (`N') is the Nobeyama data. For comparison we also show the THINGS \hi\ data as the black dashed line (labeled `T').}
\label{fig:fig4}
\end{figure*}

Therefore, we then proceed to stack \emph{all} LOS inside the  $15^{\prime\prime}$ wide tilted rings. The resulting stacked spectra are shown in Figure\,\ref{fig:fig4} for NGC\,4736 and NGC\,5055, respectively. In these figures we show the normalized stacked spectra with peak SNR higher than 5 in all data sets. The corresponding FWHM obtained for these stacked spectra are plotted in Figure\,\ref{fig:fig3} in large red symbols. The analysis of the stacked spectra (red symbols) gives the line widths for \emph{all} LOS within a radial annulus in a luminosity\,--\,weighted sense, thus also including faint LOS.

We find that the line widths of the stacked spectra including all LOS are typically larger (by 2\,--\,15\%)  than the values obtained when stacking individual high SNR LOS only. A possible explanation is that the faint LOS have systematically wider line widths than the high SNR LOS and when included in the stacks, they widen the resulting spectral line width. Since high SNR LOS are less than 20\% of the total LOS in the CARMA data (which account for $\sim$\,60\% of the interferometric flux), we cannot rely on these measurements as being representative of the underlying weaker emission.

\subsection{Comparison of FWHM from Different Instruments}
\label{r2}
\begin{figure*}[htb]
\centering
\subfiguretopcaptrue
\subfigure[hang,nooneline][Stacking all LOS]
{\epsscale{1.15}\plottwo{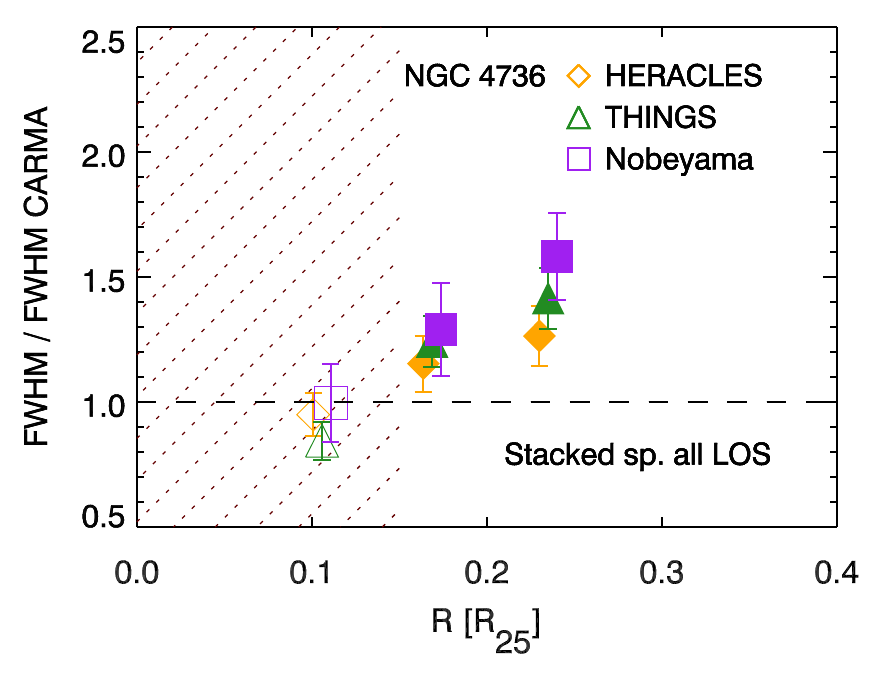}{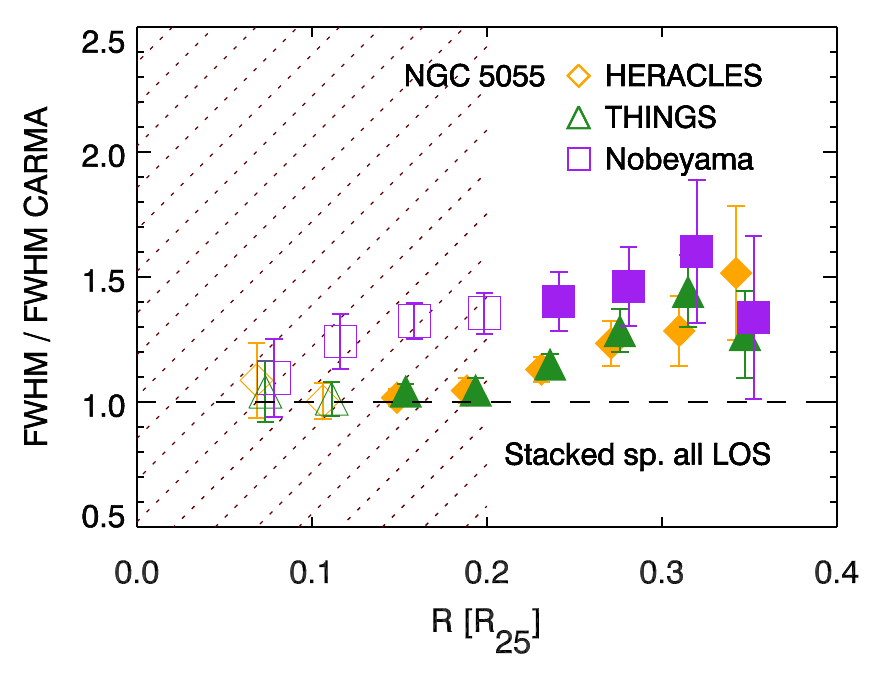}\label{fig5.a}}

\subfigure[hang,nooneline][Results involving individual LOS with SNR\,$>$\,5]
{\epsscale{1.15}\plottwo{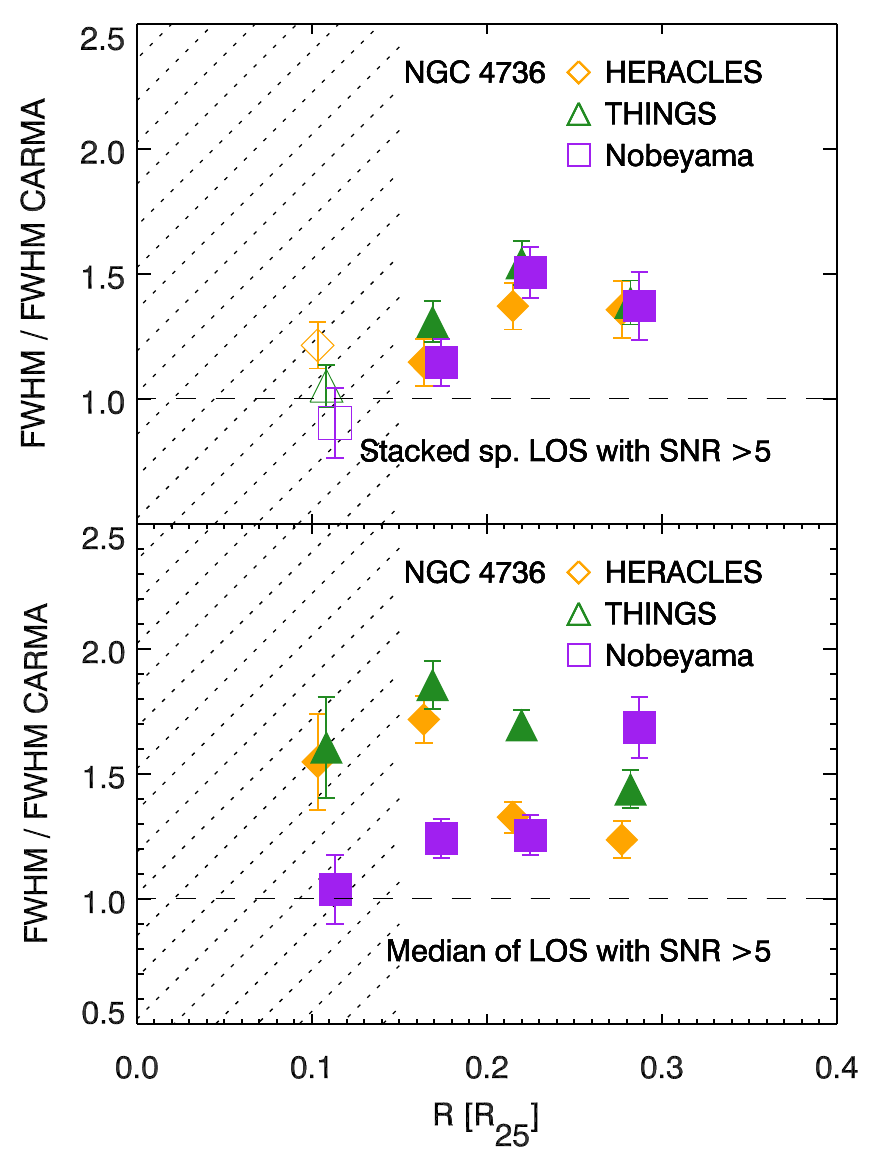}{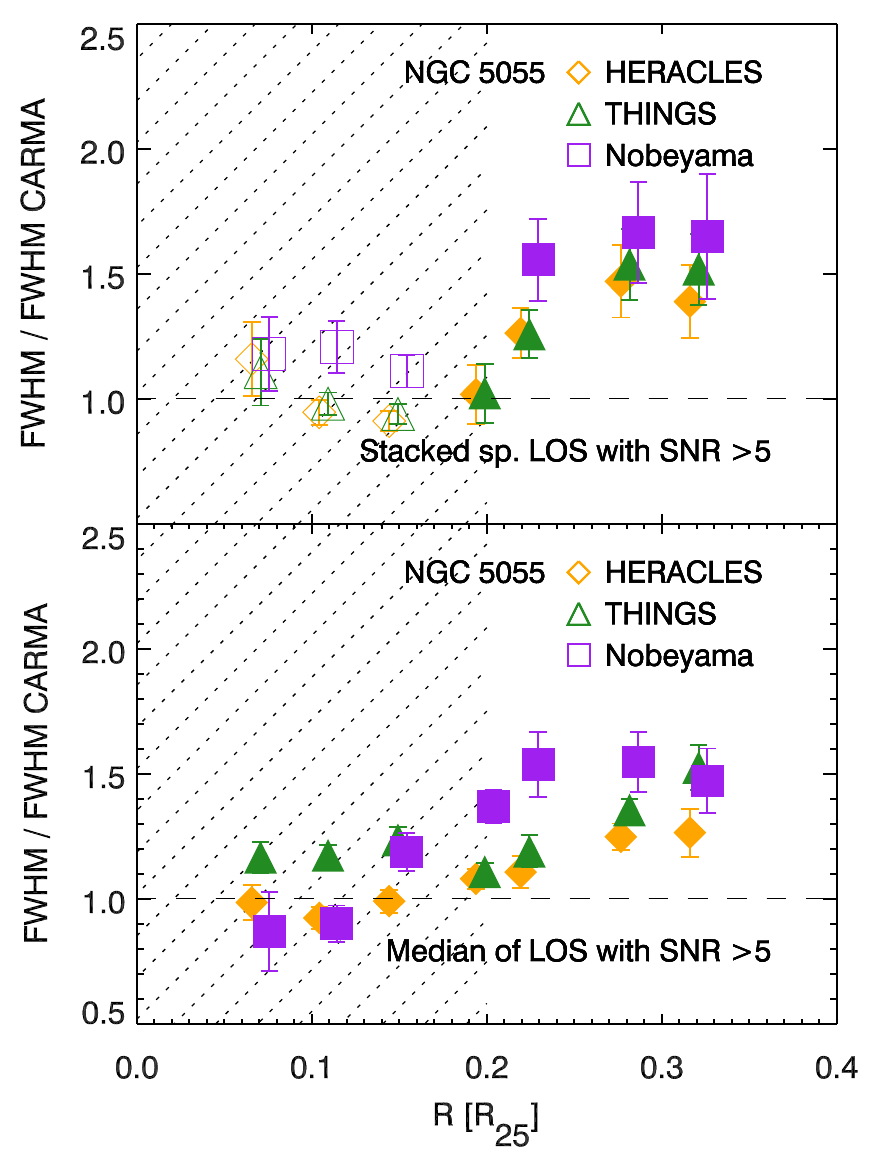}\label{fig5.b}}

 \caption{Ratio of the FWHM computed for the different instruments over the FWHM computed from the CARMA data for NGC\,4736 (left column) and NGC\,5055 (right column). \emph{Top:} Stacked spectra using all LOS inside radial bins of 15$^{\prime\prime}$. \emph{Center:} Stacked spectra of high SNR LOS. \emph{Bottom:} Median values of the high SNR individual LOS. The different symbols correspond to the different  ratios: HERACLES/CARMA in diamonds, THINGS/CARMA in triangles, and Nobeyama/CARMA in squares. The shaded areas correspond to the regions inside which beam smearing contributes more than 30\% to the measured line width measurements. The dashed line represents unity. The colors correspond to the colors in Figure\,\ref{fig:fig3}.}\label{fig:fig5}  
   \end{figure*}

A visual inspection of Figures \ref{fig:fig3} and \ref{fig:fig4} already indicates that the CARMA data give the smallest line widths of all data sets (in those areas not affected by beam smearing). Figure\,\ref{fig:fig5} shows the ratio of the FWHM line widths measured by the single\,--\,dish telescopes: HERACLES (diamonds) and Nobeyama (squares) divided by the line widths measured by CARMA for our two galaxies: NGC\,4736 (Fig.\,\ref{fig:fig5}, left column) and NGC\,5055 (Fig.\,\ref{fig:fig5}, right column). The figure is divided in two. In Fig.\,\ref{fig5.a} we show the results for the stacked spectra including all LOS. In Fig.\,\ref{fig5.b}\,we show the results involving high SNR LOS: on top the stacked spectra, and on the bottom the median values of the individual LOS. For comparison, we also show the ratio of the line widths of the THINGS \hi\ data (triangles) with the CARMA \coone\ data. Open symbols show spectra fitted by a double\,--\,horn profile (typically at small radii where beam smearing is dominant), and solid symbols correspond to Gaussian fits (typically in the galactic disk)\footnote{In Figure\,\ref{fig5.b} (bottom) we plot only filled symbols. There is no information about line profiles in this plot, since we are showing the median of individual measurements.}. The shaded regions mark the galactocentric distance where beam smearing is larger than 30\% for NGC\,4736 and NGC\,5055, respectively. Beam smearing tends to represent the large\,--\,scale gas kinematics instead of the intrinsic line profile. As this affects all data sets equally (after convolution to a common resolution), it drives any ratios of line widths toward unity, thus concealing any intrinsic differences. This is obviously the case at small radii (inside the shaded regions). At larger radii, where beam smearing is not important, the CO line widths measured from single\,--\,dish data sets (but also for the interferometric \hi\ data) are larger than the line widths measured from the interferometric CARMA data. 
 
In Table~\ref{tab2} we list the mean values for the ratios of the single\,--\,dish line widths from HERACLES and Nobemaya to the interferometric line widths from CARMA, for data only mildly affected by beam smearing,\,i.e., outside 0.15\,R$_{25}$ for NGC\,4736 and outside 0.2\,R$_{25}$ for NGC\,5055. The main finding of this paper is that the average ratio of single\,--\,dish to interferometer line width is 1.4\,$\pm$\,0.2 when looking at the \coone\, transition and taking into account only the high SNR LOS. This value is in agreement with the ratio found when taking into account all LOS (1.5\,$\pm$\,0.1). Thus single\,--\,dish observations typically trace molecular gas with larger line widths than interferometric observations. This points to the existence of a molecular gas component that is missed by the interferometer and which has larger line widths.

\subsection{Tests}\label{tests}
\subsubsection{Radial bin filling factor}\label{fillf}
As noted before, in the case of NGC\,5055 we face incompleteness for the outermost radial bins in this study (see Figure\,\ref{fig5.a}). We test how representative the measured values are, taking into account the incompleteness in these points. To do this we take the last complete radial bin in NGC\,5055, which goes from 45\,--\,60$^{\prime\prime}$  (centered at $\sim$\,0.15\,R$_{25}$).  We then take a grid of filling factors going from 90\,--\,10\% (i.e.,\, how many individual LOS within the radial bin are taken into account). We select the corresponding percentage of random individual LOS inside the radial bin and we stack them. Finally we measure the FWHM from the resulting stacked spectrum. We perform this same procedure 1000 times, in order to see how much the measurement varies by selecting random individual LOS. At the end we calculate a mean value of the FWHM computed 1000 times for each filling factor value and the dispersion among individual measurements.

We compare the ``true'' FWHM (i.e.,\,the FWHM obtained when stacking 100\% of the LOS inside the radial bin) to the mean FWHM obtained for each filling factor. For the CARMA data the variation is at most 2\% of the original value. For the HERACLES, Nobeyama, and THINGS data the variations are at most 2\%, 3\%, and 6\%, respectively. For these three datasets, the variations rise to 5\%, 12\%, and 4\%,  respectively in the case where the filling factor is of only 10\%. Looking at each data set individually we note, as expected, that the dispersion in the measurements increases with decreasing filling factor, going from $\sim$\,1\,\kmpers\ for the 90\% filling factor, up to $\sim$\,4\,\kmpers\ for the 10\% filling factor. In all cases, the dispersion is of the same order as the errors calculated for the stacked spectra ($\sim$\,2\,--\,4\,\kmpers).

\subsection{FWHM dependance on azimuthal location}
Since the circular velocities have a larger gradient along the minor axis (due to projection effects), we also test whether the angular distance to this axis affects our FWHM measurements.  For each galaxy we mask out a wedge of X degrees around the minor axis, where X ranges from 5 to 40 degrees in steps of 5 degrees. We then stack the remaining LOS and compare the FWHM to the value obtained when using no mask. For both galaxies and for all angles the differences in FWHM measurements are less than a channel width ($<$\,2.5\,\kmpers) which is 5\,--\,10\% of typical FWHM measurements, and therefore insignificant.

\subsubsection{Pointing uncertainties}
At this stage we test whether the single\,--\,dish pointing uncertainties could be responsible for the excess in the FWHM measurements. We convolve each single\,--\,dish cube with a Gaussian of width equal to the corresponding pointing uncertainty\footnote{The IRAM 30m telescope has a pointing accuracy of $\sim$\,2'' and the Nobeyama 45m telescope as a pointing accuracy of $\lesssim$\,7''. }. We compare the resulting FWHM to the FWHM measured from the original cubes. In all cases the differences between both measurements are less than 2\%. The pointing uncertainties are thus insignificant in the determination of FWHM values. 

\subsubsection{Filtering of extended emission}
In this subsection we simulate what the interferometer would detect by using the task \emph{simobserve}  in the \emph{Common Astronomy Software Applications} package (CASA) \citep{mc07}. As our galaxy template, we use the HERACLES single\,--\,dish cube, which contains all  emission (clumpy and/or diffuse). We simulate what CARMA's\,E\,--\,configuration UV\,--\,coverage would recover. We here choose CARMA's most compact configuration since this configuration recovers the largest emission scales. After constructing the interferometric cube using \emph{simobserve}, we analyze the simulated interferometric data in the same way as we did with the real data. Finally we compare the FWHM measured in the simulated interferometric observations to the HERACLES observations. We find that the HERACLES values are larger by a similar amount ($\sim$\,50\%) to the values measured in the simulated interferometric data. This further supports the idea that the information filtered by the interferometer results in narrower line widths,\,i.e.\,the extended emission filtered out by the interferometer has also larger velocity dispersion.

Finally, we caution that CARMA and Nobeyama observed the \coone\ transition whereas HERACLES observed \cotwo. The mean ratio of the Nobeyama to CARMA line widths is $\sim$\,25\% larger than the corresponding mean ratio of HERACLES to CARMA line widths. This difference might be due to different excitation temperatures within the clumpy or diffuse gas or due to different optical depth effects. With the available data we cannot differentiate between those two mechanisms.

\begin{deluxetable}{l c c} 
\tablecolumns{3}
\tablewidth{0.5\textwidth}
\tablecaption{Linewidth Ratios\label{tab2}}
\tablehead{\colhead{}  & \colhead{HERACLES$_{\cotwo}$/} & \colhead{Nobeyama$_{\coone}$/}\\
\colhead{}  & \colhead{CARMA$_{\coone}$}  & \colhead{CARMA$_{\coone}$} }
 \startdata
Stacked, all LOS \tablenotemark{a} & 1.2\,$\pm$\,0.1 & 1.5\,$\pm$\,0.1  \\
Stacked, SNR\,$>$\,5 \tablenotemark{b} & 1.3\,$\pm$\,0.1 & 1.4\,$\pm$\,0.2  \\
Median, SNR \,$>$\,5 \tablenotemark{c}& 1.3\,$\pm$\,0.2 & 1.3\,$\pm$\,0.2  
\enddata
\tablenotetext{a}{Mean of the values measured when stacking all LOS.}
\tablenotetext{b}{Mean of the values measured when stacking individual LOS with SNR \,$>$\,5.}
\tablenotetext{c}{Median of the values measured for the individual LOS with SNR \,$>$\,5.}
\end{deluxetable} 
\vspace{5mm}

\subsection{Flux Comparison}\label{flcomp}
We compare the fluxes recovered by the single\,--\,dish and by the interferometric data sets by means of integrated intensity maps (i.e., zeroth-moment maps). To construct these maps we first create  3D\,--\,masks that we use to blank the noise\,--\,dominated regions in the cubes. We construct these 3D\,--\,masks by finding regions with peak SNR larger than 5$\sigma$ in at least two consecutive channels. We then expand these masks to masks constructed in the same way, but using a SNR cut of 2$\sigma$ instead. Finally, we expand the mask by half a beam size and include adjacent velocity channels to capture all emission. We construct two masks for each galaxy; one for the interferometric CARMA data set and another one for the single\,--\,dish data sets. For the single\,--\,dish data sets we use the HERACLES cubes to construct the masks, as it has higher SNR as compared to the Nobeyama data sets. The final mask is applied to each cube and the integrated intensity is calculated. 

We compare the fluxes in three ways:
A) We compute the integrated intensity map for each data set independently (each data set with its own 3D\,--\,mask). We then measure the flux inside the region limited by  the CARMA\,--\,sensitive region (indicated in Figure\,\ref{fig:fig1}). 
B) We compute the integrated intensity map for each data cube, but use the CARMA 3D\,--\,mask in all cases.
C) We compute the integrated intensity map for each data cube, but use the HERACLES 3D\,--\,mask in all cases. We then measure the flux inside the CARMA\,--\,sensitive region. The uncertainties we state in Table\,\ref{tab3} only take into account the canonical 10\% flux calibration uncertainty for each measurement, which is likely a lower limit to the actual uncertainties.

The percentage of flux recovered by the interferometer, compared to the flux measured by the single\,--\,dish telescope, for the three different approaches are presented in Table \ref{tab3}. Method A tells us how much flux is recovered by each instrument, even if the flux is originating in not exactly the same position, but in the area where CARMA has sufficient sensitivity to pick up emission. Method B is constrained to look for emission where there is CARMA emission, whereas Method C is constrained to look for emission where there is HERACLES emission. These results show that even with Method B, where we measure flux in precisely the interferometric 3D\,--\,mask, the interferometer recovers  $\sim$\,74\,--\,81\% (depending on the galaxy) of the flux recovered by the single\,--\,dish. Using Method A, which looks for flux in the whole region were CARMA has good sensitivity, we find that in the case of NGC\,4736 the flux recovery by the interferometer is $\sim$\,52\% only, while for NGC\,5055 it is much higher (92\%). With Method C, which by construction looks for emission using the single\,--\,dish 3D\,--\,mask,  the interferometer recovers $\sim$\,35\% of the flux in the case of NGC\,4736 and $\sim$\,92\% in the case of NGC\,5055. 
In the case of NGC\,5055 the flux recovery is approximately the same, regardless of the mask used for the computation of the moment maps. In the case of NGC\,4736 the flux recovery varies from $\sim$\,35\,--\,74\%, depending on the choice of mask.

Regarding the interferometric flux measurements, we recall that during the image deconvolution process, the flux recovery will also depend on the noise properties \citep[e.g.,][]{he02}. The cleaning procedure recovers only the flux that is above a few times the noise rms. Therefore, the flux that is not being recovered by the interferometer due to the noise properties is a low amplitude component.

Finally, as a consistency check, we compare the flux from the Nobeyama 45m cubes to the flux measured by the NRAO\,12m telescope, observed as part of BIMA SONG \citep{hel03}. For NGC\,4736 we find that the flux measured in both observations agrees within 8\%, whereas for NGC\,5055 the agreement is within 15\%. In the case of NGC\,4736 the interferometer is clearly not recovering all the flux measured by the single dish. In NGC\,5055 the single\,--\,dish and interferometric flux measurements agree within their uncertainties.

\begin{deluxetable}{c c c c} 
\tablecolumns{4}
\tablewidth{0.45\textwidth}
\tablecaption{Flux Recovery by the Interferometer\tablenotemark{a}. \label{tab3}}

\tablehead{\colhead{}  & \colhead{Method A} & \colhead{Method B} & \colhead{Method C}}
 \startdata
NGC\,4736 & (52\,$\pm$\,14)\% & (74\,$\pm$\,14)\% & (35\,$\pm$\,14)\%  \\
NGC\,5055 & (92\,$\pm$\,14)\% & (81\,$\pm$\,14)\%  & (92\,$\pm$\,14)\%
\enddata
\tablenotetext{a}{The percentages show the fraction of flux recovered by the interferometer (CARMA) as compared to the single\,--\,dish (Nobeyama).The uncertainties are computed using the typical 10\% flux calibration uncertainty for each measurement.} 
\end{deluxetable} 

\section{Discussion and Summary}
\label{sum}

We present a comparison between FWHM line widths obtained from interferometric CARMA \coone\ data to single\,--\,dish \cotwo\ data from the IRAM 30m and to single\,--\,dish \coone\ Nobeyama 45m telescopes in two nearby spiral galaxies: NGC\,4736 and NGC\,5055. After convolution of the data sets to a common spatial and spectral resolution, we measure single\,--\,dish line widths that are $\sim$(40\,$\pm$\,20)\% larger than interferometric ones (when looking at the common \coone\ transition) and when taking into account the high\,--\,SNR\,--\,only LOS. If we take into account all LOS for the stacking, we get single\,--\,dish FWHM that are $\sim$(50\,$\pm$\,10)\% larger than the interferometric ones. If the interpretation of a diffuse molecular gas component is correct, we would expect the highest SNR LOS to be dominated by the emission coming from GMCs. The diffuse emission would be more prominent in lower SNR LOS. Furthermore, the flux comparisons we perform point to a flux recovery by the interferometer as low as $\sim$\,52\,--\,74\% of the flux measured by the single\,--\,dish telescope over the same region in the case of NGC\,4736. This result is consistent to what \citet{pe13} found in the case of M51. In the case of NGC\,5055 the interferometric flux recovery is higher $\sim$\,81--\,92\%, and taking into account the uncertainties in the flux determination, we cannot rule out a full flux recovery by the interferometer for this galaxy. In summary, we measure large line widths and larger flux recoveries from the single\,--\,dish observations as compared to the interferometric ones. The molecular gas `unseen' by the interferometer could have two possible configurations.

One possibility is that this gas is diffuse,\,i.e., with lower average densities as compared to the gas present inside GMCs. There is evidence for diffuse molecular gas in our Galaxy observed through absorption lines \citep[][and references therein]{sn06, bu10}. The column densities of the diffuse gas are found to be $\lesssim$\,30 times smaller than those in molecular clouds and with typical temperatures determined from diffuse H$_2$  to be $\gtrsim$\,30\,K \citep{li10}. This diffuse gas would need to extend over volumes larger than the maximal scales traced by the interferometer, and thus be filtered out. The other possibility is that the gas is inside small molecular clouds which are separated among each other by less than a synthesized beam size. This configuration would appear as a large regular structure to the interferometer, and therefore would be filtered out. The physical properties of these clouds (temperature and density) may be (or not) comparable to the values measured inside GMCs. In this case larger velocity dispersion would be the result of cloud\,--\,to\,--\,cloud motions. With the available data we cannot distinguish between these two possible scenarios. Detailed analysis of CO isotopes might help discriminate between diffuse and dense gas \citep{li10}, as would be higher spatial resolution observations.

In Galactic studies, the existence of a diffuse molecular gas is well accepted \citep[e.g.,\,][]{li12, sh08}. However, these studies are hampered by uncertain distances and projection effects that make it  difficult to quantify the amount of gas present in this pervasive component. Studies of nearby galaxies offer an `outside view' that do not face these difficulties, though measurements are still sparse. Firmly establishing the existence of such a spatially extended and high\,--\,velocity dispersion molecular gas phase would have important implications. The velocity dispersion measured from molecular gas emission (thought to arise from GMCs only)  has been broadly used for molecular gas scaling relations, or Larson's laws \citep{la81}, among other things, to study star formation. The presence of two different molecular gas configurations could imply that these scaling relations may not be constant throughout the galaxy  (depending on the dominant molecular gas configuration). Specifically in the extragalactic context, where different instruments (single\,--\,dish and interferometer) yield different velocity dispersion measurements, studies of a larger sample are needed in order to quantify how the scaling relations (for example velocity dispersion\,--\,molecular cloud size) would be affected.

As a final comment, a complimentary observational approach to address the same question is to study edge\,--\,on galaxies, as was done by \citet{ga92}. They detect molecular gas up to $\sim$\,1\,kpc off the plane of the galaxy. They interpret this as evidence of a thick molecular disk in NGC\,891, which is also in agreement with a diffuse molecular gas component, since higher velocity dispersions in principle imply a larger scale height of the molecular gas disk. Future observations with ALMA, using both interferometric  observations as well as total power measurements, will be crucial to shed more light on the nature of this `diffuse' molecular gas phase in galaxies.

\acknowledgements
We thank the anonymous referee for very thoughtful and useful comments. ACP appreciates interesting discussions with J\'er\^ome Pety on flux measurements. ACP also acknowledges support from the DFG priority program 1573 `The physics of the interstellar medium', from the IMPRS for  Astronomy \& Cosmic Physics at the University of Heidelberg, and from the Mexican National Council for Science and Technology (CONACYT).   A.D.B. wishes to acknowledge partial support from the CAREER grant NSF AST-0955836 and an Alexander von Humboldt Foundation Fellowship.

\bibliography{mybib}
\end{document}